\documentclass[aps,prd,showpacs,nofootinbib,superscriptaddress,preprintnumbers,twocolumn]{revtex4-1}
\usepackage{graphicx,amsmath,amssymb,soul,bbm,dcolumn,booktabs, multirow,array,float,enumitem,lmodern,dsfont,microtype,nicefrac,rotating,adjustbox,bm,xparse}
\usepackage[dvipsnames]{xcolor}
\usepackage[utf8]{inputenc}
\usepackage{upgreek}
\AtBeginDocument{
\heavyrulewidth=.08em
\lightrulewidth=.05em
\cmidrulewidth=.03em
\belowrulesep=.65ex
\belowbottomsep=0pt
\aboverulesep=.4ex
\abovetopsep=0pt
\cmidrulesep=\doublerulesep
\cmidrulekern=.5em
\defaultaddspace=.5em
}

\newcolumntype{C}{>{$}c<{$}}
\AtBeginDocument{
\heavyrulewidth=.08em
\lightrulewidth=.05em
\cmidrulewidth=.03em
\belowrulesep=.65ex
\belowbottomsep=0pt
\aboverulesep=.4ex
\abovetopsep=0pt
\cmidrulesep=\doublerulesep
\cmidrulekern=.5em
\defaultaddspace=.5em
}
\newcommand{\be}{\begin{eqnarray}}
\newcommand{\ee}{\end{eqnarray}}

\def\fm {\mathop{\hbox{fm}}}
\def\MeV {\mathop{\hbox{MeV}}}

\def\beq{\begin{equation}}
\def\eeq{\end{equation}}
\def\beqs#1\eeqs{\beq\begin{split} #1 \end{split}\eeq}

\def\comment#1{}

\def\ket#1{\left| #1 \right\rangle}

\def\opbraket#1#2#3{ \left\langle #1 \left| #2 \right| #3 \right\rangle}

\newcolumntype{L}{>{$}l<{$}} 
\newcolumntype{S}{>{\footnotesize $}l<{$\normalsize}} 

\def\*#1{\mathbf{#1}}

\newcolumntype{R}[2]{%
    >{\adjustbox{angle=#1,lap=\width-(#2)}\bgroup}%
    c%
    <{\egroup}%
}
\newcommand*\rot{\multicolumn{1}{R{75}{1em}}}

\usepackage{array}

\usepackage[colorlinks=true,backref=false, linktocpage=true,
citecolor=PineGreen,urlcolor=PineGreen,linkcolor=PineGreen,pdfpagemode=UseOutlines]{hyperref}
\hypersetup{%
  bookmarksnumbered=true,
  pdftitle = {},
  pdfsubject = {},
  pdfauthor = {},
  pdfkeywords = {}
}

\begin{document}
\title{Three pion spectrum in the \texorpdfstring{$I=3$}{} channel from lattice QCD}

\author{C.\ Culver}
\email{chrisculver@email.gwu.edu}
\affiliation{The George Washington University, Washington, DC 20052, USA}
\author{M.\ Mai}
\email{maximmai@gwu.edu}
\affiliation{The George Washington University, Washington, DC 20052, USA}
\author{R.\ Brett}
\email{rbrett@gwu.edu}
\affiliation{The George Washington University, Washington, DC 20052, USA}
\author{A.\ Alexandru}
\email{aalexan@gwu.edu}
\affiliation{The George Washington University, Washington, DC 20052, USA}
\affiliation{Department of Physics, University of Maryland, College Park, MD 20742, USA}
\author{M.\ D\"oring}
\email{doring@gwu.edu}
\affiliation{The George Washington University, Washington, DC 20052, USA}
\affiliation{Thomas Jefferson National Accelerator Facility, Newport News, VA 23606, USA}
%

\begin{abstract}
Three-body states are critical to the dynamics of many hadronic resonances.  We show that lattice QCD calculations have reached a stage where these states can be accurately resolved. We perform a calculation over a wide range of parameters and find all states below inelastic threshold agree with predictions from a state-of-the-art phenomenological formalism. This also illustrates the reliability of the formalism used to connect lattice QCD results to infinite volume physics. Our calculation is performed using three positively charged pions, with different lattice geometries and quark masses.
\end{abstract}

\maketitle

\section{Introduction}

It is important to understand hadron interactions in terms of quark-gluon dynamics as they emerge from QCD. Three-hadron systems present the next hurdle in this quest, primarily for resonances with final state decays to three particles.  One such example is the Roper resonance $N(1440)1/2^+$ which couples strongly to $\pi\pi N$.  The analysis of experimental data exposes the complex analytic structure of the resonance~\cite{Arndt:2006bf,Doring:2009yv}.  Lattice studies to date have difficulty finding a finite volume state along the Roper mass trajectory~\cite{Virgili:2019shg,Liu:2016rwa, Sun:2019aem, Lang:2016hnn}.  Note that none of the studies in the Roper channel include three hadron interpolators, in contrast with this work. Jefferson Lab, ELSA, MAMI and other facilities have experimental programs dedicated to studying excited baryons~\cite{Crede:2013sze, Aznauryan:2012ba, Klempt:2009pi} making the need for three body analyses critical.  Another such example is the $a_1(1260)$ resonance, which decays to $\rho\pi$ and $\sigma\pi$ intermediate states before its final state of three pions.  Moreover, exotic mesons with quantum numbers forbidden by quark models need to be understood.  With ongoing experiments, e.g.~GlueX at Jefferson Lab, searching for exotic states, there is a demand for theoretical determinations of the QCD spectrum below and in the region where such exotic states may lie.

Lattice QCD (LQCD) is used to determine hadron properties and interactions as they arise from the quark-gluon dynamics providing complementary information to experimental data (for example, in LQCD we can freely modify the quark masses and number of flavors). In this approach hadron interactions are probed through the spectrum of interacting (many) hadron states in finite volume.  The finite-volume spectrum has to be connected to infinite-volume scattering information through the use of quantization conditions~\cite{Luscher:1990ux, Luscher:1986pf}.  This formalism has been used extensively in the two-hadron sector in the last decade, see for example Refs.~\cite{Pelissier:2012pi, Guo:2016zos,Guo:2018zss,Culver:2019qtx,Mai:2019pqr,Bulava:2016mks,Brett:2018jqw,Andersen:2018mau,Alexandrou:2017mpi,Bali:2015gji,Feng:2014gba,Feng:2010es,Orginos:2015aya,Beane:2011sc,Aoki:2011yj,Dudek:2012xn,Dudek:2012gj,Mohler:2013rwa,Prelovsek:2013ela,Mohler:2012na,Lang:2012sv,Lang:2011mn,Guo:2016zos,Helmes:2017smr,Liu:2016cba,Helmes:2015gla,Wilson:2014cna,Wilson:2015dqa,Moir:2016srx}, and Ref.~\cite{Briceno:2017max} for a review.  Lattice calculations in the three-hadron sector have focused on the lowest states~\cite{Beane:2007es,Detmold:2008fn,Detmold:2008yn} and only recently pioneering calculations of the higher levels have been reported~\cite{Horz:2019rrn,Woss:2019hse}. In parallel, a significant theoretical effort was dedicated to developing the formalism to connect the three-body finite volume data to infinite-volume amplitudes~\cite{Mai:2019fba, Blanton:2019vdk, Mai:2018djl, Guo:2019hih,Romero-Lopez:2019qrt,Mai:2018djl,Zhu:2019dho,Guo:2018ibd, Doring:2018xxx,Guo:2018xbv,Romero-Lopez:2018rcb,Klos:2018sen, Mai:2017bge,Guo:2017crd, Guo:2017ism, Hammer:2017kms,Briceno:2018aml,Briceno:2018mlh,Briceno:2017tce,Guo:2016fgl,Hansen:2016fzj,Hansen:2015zga,Jansen:2015lha,Hansen:2014eka,Polejaeva:2012ut,Roca:2012rx,Briceno:2012rv,Bour:2012hn, Kreuzer:2012sr,Kreuzer:2009jp,Kreuzer:2008bi, Meng:2017jgx, Hammer:2017uqm, Meissner:2014dea, Bour:2011ef, Kreuzer:2010ti}, see Ref.~\cite{Hansen:2019nir} for a review.

Here we show that LQCD results for three-pion energies agree very well with theoretical expectations, over a wide set of parameters. Using multiple volumes and quark masses we compute a large number of energy levels in the elastic three-pion scattering region. We compare them to predictions from chiral perturbation theory inspired models, connected to finite volume through relativistic three-body quantization conditions (R3Q) preserving unitarity~\cite{Mai:2019fba, Mai:2019pqr, Mai:2018djl, Doring:2018xxx, Mai:2017bge}.  We find excellent agreement indicating that both approaches have reached maturity. This paves the way to attack more challenging problems, such as coupled-channel extensions needed for the realistic description of Roper, $a_1(1260)$, exotics and other hadrons.

The results presented here are for three-pion states in maximal isospin ($I=3$) in QCD with two mass-degenerate quark flavors, for two different quark masses, corresponding to pion masses of $220\MeV$ and $315\MeV$.
For each quark mass the calculation is done using three different geometries, one cubic and two different elongations in a single spatial direction.  We use elongations as a cost effective way to map out the elastic scattering region.
The outline of the presentation is as follows. First we highlight some important details of the lattice calculation.  Then the R3Q formalism is presented, including an extension required to accommodate elongated boxes.  Finally, the results for the finite-volume spectrum, as determined from LQCD and R3Q predictions, are presented.

\section{Lattice Methods}\label{sec:lattice}

\begin{table}[t]   
\caption{Details of the $N_f=2$ ensembles used in this study.  $\eta$ is the elongation, $a$ is the lattice spacing.  $N_{\text{cfg}}$ is the number of monte carlo configurations for each ensemble.}
\label{table:gwu_lattice}  
\begin{ruledtabular}
\begin{tabular}{@{}*{13}{>{$}l<{$}}@{}}                                                     
\text{Label}~& N_t\times N_{x,y}^2\times N_z & ~\eta~ & ~a[\fm]~   & ~N_\text{cfg}~  & ~am_{\pi}~ \\
\midrule                                                                                               
\mathcal{E}_1&48\times24^2\times24  &  1.00  & 0.1210(2)(24) & 300   & 0.1931(4)\\ 
\mathcal{E}_2&48\times24^2\times30  &  1.25 & -        & -     &    0.1944(3)\\
\mathcal{E}_3&48\times24^2\times48  &  2.00  & -        & -     &    0.1932(3)\\
\mathcal{E}_4&64\times24^2\times24  &  1.00  & 0.1215(3)(24)& 400  & 0.1378(6) \\
\mathcal{E}_5&64\times24^2\times28  &  1.17 &      -   & -     & 0.1374(5)\\
\mathcal{E}_6&64\times24^2\times32 &  1.33 &      -   & -     & 0.1380(5)\\
\end{tabular}  
\end{ruledtabular}
\end{table} 

In infinite volume hadron interactions are encoded using scattering amplitudes that are decomposed into distinct partial-waves using rotational symmetry.  LQCD can access scattering information only indirectly via the discrete spectrum of hadron states in finite-volume.  Analogous to infinite-volume states having definite angular momentum, finite-volume states are labeled by irreducible representations (irreps) of the reduced rotational symmetry groups.  Moreover, the reduced rotational symmetry mixes partial waves, complicating the mapping between finite and infinite volume.  

To extract the finite volume spectrum of three pions we use cubic lattices as well as lattices with an elongation in one spatial direction.  The momenta on a periodic lattice are quantized in units of $2\pi/L$, where $L$ is the length of the lattice.  By increasing the length in a single direction we lower the momenta and thus the energy of multi-hadron states where the constituent hadrons have non-zero momenta.  For example, in our largest elongation (${\cal E}_3$), in the $A_{1u}$ irrep there are seven energy levels below the $5m_{\pi}$ threshold whereas the cubic box has only two.  The details of the ensembles are found in Table~\ref{table:gwu_lattice}.  

For cubic lattices the infinite-volume angular momentum symmetry group $SO(3)$ is reduced to $O_h$.  The elongation further reduces the symmetry group from $O_h$ to $D_{4h}$. For boosted systems, where the total momentum vector $\*P$ is aligned along the elongation axis, the relevant symmetry group is the little group that leaves such momentum vectors invariant, $C_{4v}$. While other boosts are possible, they will not be considered in this study. 
The mapping between angular momentum $\ell$ and the irreps of these symmetry groups is described in Ref.~\cite{Lee:2017igf}.


The finite-volume spectrum is determined by fitting the large time behaviour of temporal correlation functions between hadronic operators.  These operators
are constructed to increase the overlap with the states of interest.  We use operators constructed from products of single pion interpolators of the form
\beq
\pi^+(\Gamma(\*p),t)=\bar{d}(t)\Gamma(\*p)u(t),
\label{equation:pion}
\eeq
where the $u/d$ quark fields are vectors in position, spin, and color space, and the momentum matrix $\Gamma(\*p)=e^{i\*p\cdot\*x}\gamma_5$ is a matrix in just position and spin space.
Three-pion operators are projected onto row $\lambda$ of the irrep $\Upgamma$ of a group $G$ using 
\beqs
 {\cal O}_{\pi_1\pi_2\pi_3}=\sum_{g\in G}&U^{\Upgamma}_{\lambda\lambda}(g)\text{det}(R(g)) \\
 \times\,&\pi^+(R(g)\*p_1)\pi^+(R(g)\*p_2)\pi^+(R(g)\*p_3),
 \label{equation:project_op}
\eeqs
where $U^\Upgamma_{\lambda\lambda}(g)$ is the representation matrix of the group element $g$ in row $\lambda$ of the irrep $\Upgamma$, and
$R(g)$ is the rotation corresponding to $g$. Explicit projection coefficients for the operators used are listed in Appendix~\ref{appendix:operators}. Note that while a two pion state in maximal isospin is restricted to even total parity, the same is not true of a three pion state, and we will study irreps with both even and odd total parity.

With a large basis of such interpolating operators, we perform a variational analysis on a matrix of correlation functions and extract the stationary state energies from a generalized eigenvalue problem (GEVP)~\cite{Luscher:1990ck,Michael:1982gb,Blossier:2009kd}. Due to the high statistical precision with which we are able to estimate the temporal correlation functions, we must take extra care when fitting the GEVP eigenvalues.   We account for contamination from excited states, and time dependent wraparound effects due to the finite temporal extent of the lattice.  To extract energy levels we perform three exponential fits, discussed further in Appendix~\ref{appendix:fitting}.

The first step in computing correlation functions is to perform the Wick contractions between creation and annihilation operators. The resulting correlation function requires the computation of expensive all-to-all quark propagators. To compute them we use Laplacian-Heaviside smearing~\cite{Peardon:2009gh}. This allows us to factorize the correlation functions in terms of products of the form $\Gamma(\*p)\tilde{M}^{-1}(t,t_f)$, where $\tilde{M}^{-1}$ is a quark propagator.  These \textit{quark lines} can be precomputed for reuse in multiple elements of the correlation matrix. Further numerical speedup is obtained by using common subexpression elimination as in Ref.~\cite{Horz:2019rrn}.

\section{Three-body Finite Volume Unitarity}\label{sec:threeBody}

The minimal set of kinematic variables required to describe relativistic elastic three-to-three scattering spans an 8-dimensional space. A convenient parametrization in terms of a one-particle spectator and two-body sub-channels, each represented by a tower of partial waves, was derived in Ref.~\cite{Mai:2017vot}. The discretized version of this relativistic, unitary approach~\cite{Mai:2017bge} is the basis of the finite-volume analysis presented in this section.

The details of this relativistic 3-body quantization condition and its implementation for systems with arbitrary boosts, and irreps can be found in Refs.~\cite{Mai:2017bge,Mai:2018xwa,Mai:2019fba, Doring:2018xxx}. The condition for finding an interacting energy eigenvalue $E^*_{\text{cm}}$ is
\begin{align}
    \infty &= \sum_{
    \langle {\* p}\rangle,
    \langle {\* q}\rangle}
    c^\Upgamma_{\langle \* p \rangle}
    c^\Upgamma_{\langle \* q\rangle}
    \\\nonumber
    \times&
    \Big\langle
    v_{\*p_2,\*p_3}
    \Big[
    B(E^*_{\text{cm}})
    +
    E_{L\eta}
    ~
    \tau_{L\eta \* P}^{-1}(E^*_{\text{cm}})
    \Big]_{\* p_1, \* q_1}^{-1}
    v_{\*q_2,\*q_3}
    \Big\rangle_{\substack{
    {\* p_i}\in {\langle \* p\rangle} \\
    {\* q_j}\in {\langle \* q\rangle}
    }}\,,
\end{align}
where $c^\Upgamma_{\langle \* p \rangle}$ are projection coefficients to an irrep $\Upgamma$ for the set of pion three-momenta $\langle \* p \rangle$ (the values of these coefficients are tabulated in Appendix~\ref{appendix:operators}.) The object in the square brackets is a matrix in the space of in/outgoing spectator momenta. Angled brackets denote symmetrization with respect to the three-set of in/outgoing three-momenta. $B$ and $\tau$ describe the two-to-one and two-pion (sub-channel) interactions, respectively, and $[E_{L\eta}]_{\* p \* q}=\delta_{\* p \* q}2L^3\eta\sqrt{m_\pi^2+{\* p}^2}$
where $\eta$ is the elongation, see Table~\ref{table:gwu_lattice}. The two-pion interaction is parametrized by a tower of functions $\tau$ in angular momentum $\ell$ coupled to the asymptotic states via a function $v$ that is free of right-hand cuts. Due to the suppression of higher partial waves ($\ell>0$) seen in both phenomenological and lattice studies~\cite{Mai:2019pqr,Horz:2019rrn,Bulava:2016mks,Dudek:2012gj} we neglect them. Additionally, since $\tau$ is defined in the rest-frame, it explicitly depends on the total momentum of the three-body system $\* P$.

Besides external parameters $L$, $\eta$, and $\* P$, there are two places where  parameters determining the dynamics enter the above quantization condition. First, the term $B$ corresponds to one-pion exchange and the genuine three-body force $C$. Second, the two-pion interaction encoded in the diagonal matrix $\tau$ is chosen to agree with the modified inverse amplitude method~\cite{Truong:1988zp,Pelaez:2006nj}, see also Refs.~\cite{Mai:2018xwa,Mai:2019fba,Mai:2017vot} for details on implementation. This method depends on four low-energy constants $\{l_{1}^r,l_{2}^r,l_{3}^r,l_{4}^r\}$ for a fixed regularization scale $\mu=770$~MeV.

We use the quantization conditions for predictions of the finite-volume energy eigenvalues. Following the procedure from previous studies~\cite{Mai:2018xwa,Mai:2019fba,Mai:2017vot}, we fix $C=0$.  We do not fit the LECs here, but use the values determined elsewhere.  To assess the uncertainties in our predictions we use two different LECs: $\{-6.032,+5.455,+0.816,+5.600\}\cdot10^{-3}$ from Ref.~\cite{Gasser:1983yg} and $\{+11.625, -0.695, +0.008,+52.411\}\cdot10^{-3}$ from a recent lattice-driven determination in Ref.~\cite{Culver:2019qtx}, denoted by GL and GW, respectively. The GW set provides the most consistent predictions, as it is determined from a fit to two-pion energy spectra on the same set of ensembles of Table~\ref{table:gwu_lattice}. The resulting predictions are collected in Appendix~\ref{appendix:energies} and shown in Fig.~\ref{fig:spectrum}.

\section{Results}\label{sec:results}

\begin{figure*}[t]
    \centering
     \includegraphics[width=\linewidth, trim = 0.1cm 1cm 2.2cm 0.7cm]{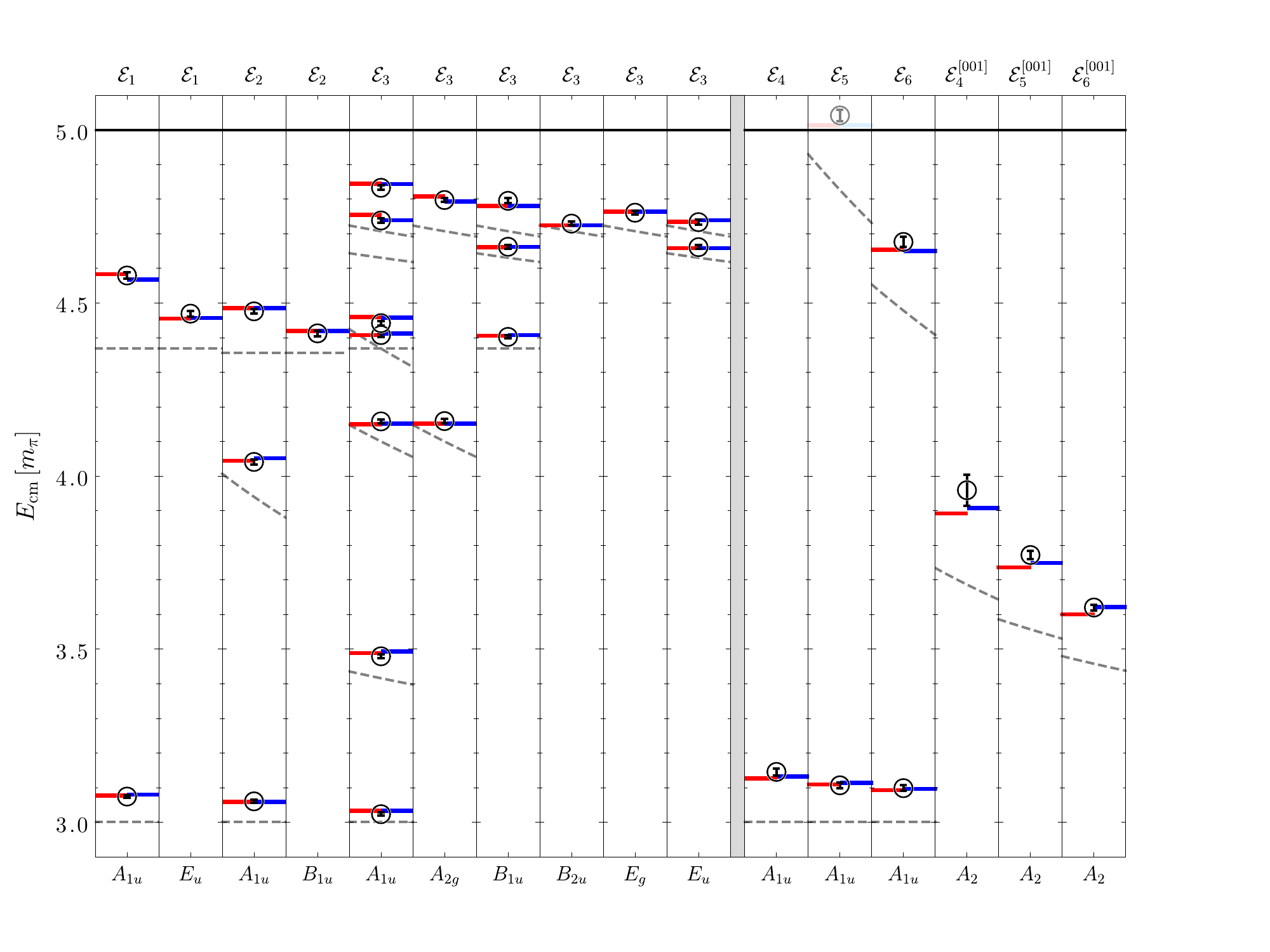}
     \caption{Finite-volume centre-of-mass energies, in units of $m_{\pi}$, for three pions in maximal isospin at two different pion masses: $\mathcal{E}_{1,2,3}$ for $315\MeV$ and $\mathcal{E}_{4,5,6}$ for $220\MeV$ (separated by the gray column).  For each pion mass there is one cubic box (${\cal E}_{1,4}$) and two elongated boxes (${\cal E}_{2,3,5,6}$).  The separate columns distinguish different irreps of the rotational symmetry group containing energies below the inelastic threshold, $5m_\pi$ (solid black line).  The parity of the irreps is specified by $g$ (even) and $u$ (odd).  The data points are the LQCD energy levels with error bars inside of the circles.  The red (left) and blue (right) solid lines in each column are the predictions from R3Q using GL or GW LECs, respectively. The dashed lines are the non-interacting energy levels. We plot them as a function of $\eta$ to distinguish the levels that depend on the elongation.  Boosted frames with non-zero total momentum are denoted by the superscript $[001]$ indicating a single unit of momentum in the elongation ($z$) direction.
     }
\label{fig:spectrum}
\end{figure*}

The three-pion ($I=3$) finite-volume spectra are shown in Fig.~\ref{fig:spectrum}, together with the non-interacting energy levels, and the predicted energy levels from R3Q. In total we extract 30 energy levels below $5m_\pi$ across the six ensembles listed in Table~\ref{table:gwu_lattice}. Precise values for lattice and R3Q levels are tabulated in Appendix~\ref{appendix:energies}.

As the elongation $\eta$ is increased, naturally the spectrum becomes much denser. This is most striking when considering the $A_{1u}$ irrep for the heavier quark mass.  At $\eta=1$ (i.e.~the cubic volume ${\cal E}_1$) we find two energies below threshold, in contrast to finding seven energies below the inelastic threshold at our largest elongation, $\eta=2$ (${\cal E}_3$). In a similar vein, as we increase $\eta$ we find energies appearing below the inelastic threshold in irreps where, at smaller elongations, no energy levels exist.

Comparing the lattice and predicted spectra in Fig.~\ref{fig:spectrum}, we find good agreement for both sets of LECs.  Energies belonging to different irreps are dominated by different partial waves.  For some levels one partial wave is dominant; as an example consider the levels between $4.4m_{\pi}$ and $4.6m_{\pi}$ in the first two columns of Fig.~\ref{fig:spectrum}.  The lowest partial wave contribution to the $A_{1u}$ level is from $S$-wave, while the $E_u$ level is dominated by $D$-wave contributions (no $S$-wave mixing). We remind the reader that only $S$-wave contributions are considered in the $\pi\pi$ interaction, while the isobar-spectator interaction encoded in the $B$-term (required by unitarity) contains higher partial waves. For further details see the discussion in Ref.~\cite{Mai:2019fba}. For other levels, the mixing of partial waves is important.  The R3Q predictions reproduce this pattern in all irreps, even for the elongated lattices, where angular momentum mixing is more severe.  Note that this mixing is not explicitly imposed in R3Q, showing that the $S$-matrix principle of unitarity combined with two-pion scattering data, post-dicts the same splitting of finite-volume levels.

We compare the predictions generated by GL or GW LECs
by computing their corresponding correlated $\chi^2/n$ where $n$ is the number of data points. 
A $\chi^2/n\approx 1$ would indicate that the LECs and three-body force used provide a good description of the data without fitting.
When using the GW LECs~\cite{Mai:2019pqr}, and including all lattice energies below $5m_{\pi}$, the reduced $\chi^2/n$ is 2.68. Excluding correlations reduces this value to 1.90.  Using the GL LECs~\cite{Gasser:1983yg}, the reduced $\chi^2/n$ is 4.86 and 2.23, with and without correlations, respectively.  As expected, the GW LECs produce predictions in better agreement with the lattice data.  Examining the $\chi^2/n$ with GW LECs on the $315\MeV$ ($\mathcal{E}_1$,$\mathcal{E}_2$,$\mathcal{E}_3$) and $220\MeV$ (${\cal E}_4$, ${\cal E}_5$, ${\cal E}_6$) ensembles independently gives 2.93 and 1.86, respectively.  This is unsurprising as chiral perturbation theory is more reliable at lower pion masses.  Since the LECs were not fitted to this data, and the contact term is set to zero, there is a tension between predictions and lattice data.  Preliminary work in this area, in which we only vary the three-body contact term, indicates that we will be able to constrain the three body force.

\section{Conclusion}\label{sec:summary}

We compute the three pion ($I=3$) finite volume spectrum at two different quark masses using LQCD.  We make use of lattices elongated in a single spatial direction, and boosts along the same axis, to capture additional states below the inelastic threshold.  In total 30 energy eigenvalues are extracted within the elastic scattering region, 23 of them at a pion mass of $315\MeV$, the remaining seven at $220\MeV$.

The lattice results agree with predictions of the finite volume spectrum from a state-of-the-art three body relativistic unitary finite-volume formalism, extended to accommodate elongations. The physical input to the formalism is a three-body contact term and the two-pion interaction.  We set the former to zero, and parametrize the latter by the modified inverse amplitude method with two sets of LECs, one from a fit to experimental data and another one determined from lattice calculations of pion-pion scattering on the same ensembles.  Not surprisingly, the LECs determined from lattice calculations provide the better prediction.  There is some tension between the lattice data and prediction, which may be due to the three-body term being set to zero.  Future work in fitting, allowing the LECs to vary and the inclusion of pion mass correlations, will reduce this tension.

Success here shows that both lattice and phenomenological efforts reached maturity and can be used to constrain three-body physics in QCD, for example the Roper and $a_1(1260)$ resonances.  

\bigskip

\begin{acknowledgments}
CC, RB, and AA are supported in part by U.S. DOE Grant No. DE-FG02-95ER40907. MM and MD are supported  by the National Science Foundation CAREER grant PHY-1452055 and by the U.S.  Department of  Energy, Office of Science, Office of Nuclear Physics under contract no. DE-AC05-06OR23177.
\end{acknowledgments}

\bibliography{ALL-REF.bib}

\appendix
\begin{widetext}
\newpage

\section{Temporal Wrap-Around Effects}\label{appendix:fitting}
Since we extract the finite-volume spectrum by examining the large time behaviour of temporal correlation functions,
care must be taken when deducing appropriate fit forms. In particular, contributions due to the finite temporal extent of the lattice become significant when considering complicated multi-hadron interactions, coupled with the high statistical precision available in this study. As lattice QCD calculations are carried out in imaginary time with a finite temporal extent, it is useful to examine three-pion correlation functions in the following form
\beq
C_{ij}(t) = \frac{1}{Z(T)} \sum_{n,m} e^{-E_n(T-t)} e^{-E_mt}
\opbraket{n}{{\cal O}_i^{3\pi}}{m} \opbraket{m}{{\cal O}_j^{3\pi}}{n},
\eeq
where ${\cal O}^{3\pi}$ is some three-pion interpolating operator, as described in the main text. For large temporal extent $T$, contributions from $\ket{n}=\ket{0}$ (i.e.~the state of interest) will dominate. Contributions from $\ket{n}\neq\ket{0}$ will appear solely due to the finite temporal extent and are often referred to as ``thermal states''.

The next most significant non-zero term appears when $\ket{n}=\ket{\pi}$, with the leading contribution containing terms proportional to $e^{-E_{\pi}(T-t)} e^{-E_{2\pi}t}$, which is time dependent. We must account for this time dependent ``temporal wraparound'' explicitly by fitting the correlation functions to three exponentials: one for the state of interest, one for residual early time excited state contamination present in the GEVP, and one to capture this leading wraparound term.  
This methodology was employed by the HadSpec collaboration~\cite{Dudek:2012gj}. They propose modifying the correlation matrix itself to account for thermal corrections and then do a regular fit for the eigenvalues. In our testing for a previous study~\cite{Culver:2019qtx} we found that we get the same results when introducing thermal corrections as additional fit parameters in the eigenvalue fit for an unmodified correlation matrix. This allows different thermal corrections for each energy level. Note that we use as fit parameters both the decay rate and spectral weight for the thermal correction. Our results are consistent with the expected decay rates for thermal corrections. For the example discussed below the lowest channel has $\delta E\approx E_{2\pi}-E_\pi$, and the spectral weights are indeed small since it is proportional to $e^{- E_\pi T}$.

An example of this fitting using the ansatz $a_1 e^{-b_1 t} + a_2e^{-b_2 t} + a_3e^{-b_3 t}$, where $\{a_i,b_i\}$ are fit parameters, to describe the correlation functions is shown in Fig.~\ref{fig:corr_fit}.  The effective masses of the ground state, and first excited state energy levels extracted in the $A_{1u}$ irrep for ensemble $\mathcal{E}_1$ are shown with best-fit curves overlaid. The resulting fit parameters can be found in Table~\ref{table:fit_params}.  Note that the parameters are ordered such that $i=1$ represents the excited state, $i=2$ the state of interest, and $i=3$ the thermal state.  We find that both single- and two-exponential fit forms are insufficient to describe these correlation functions.  

\begin{figure}[H]
    \centering
    \includegraphics[width=0.75\linewidth]{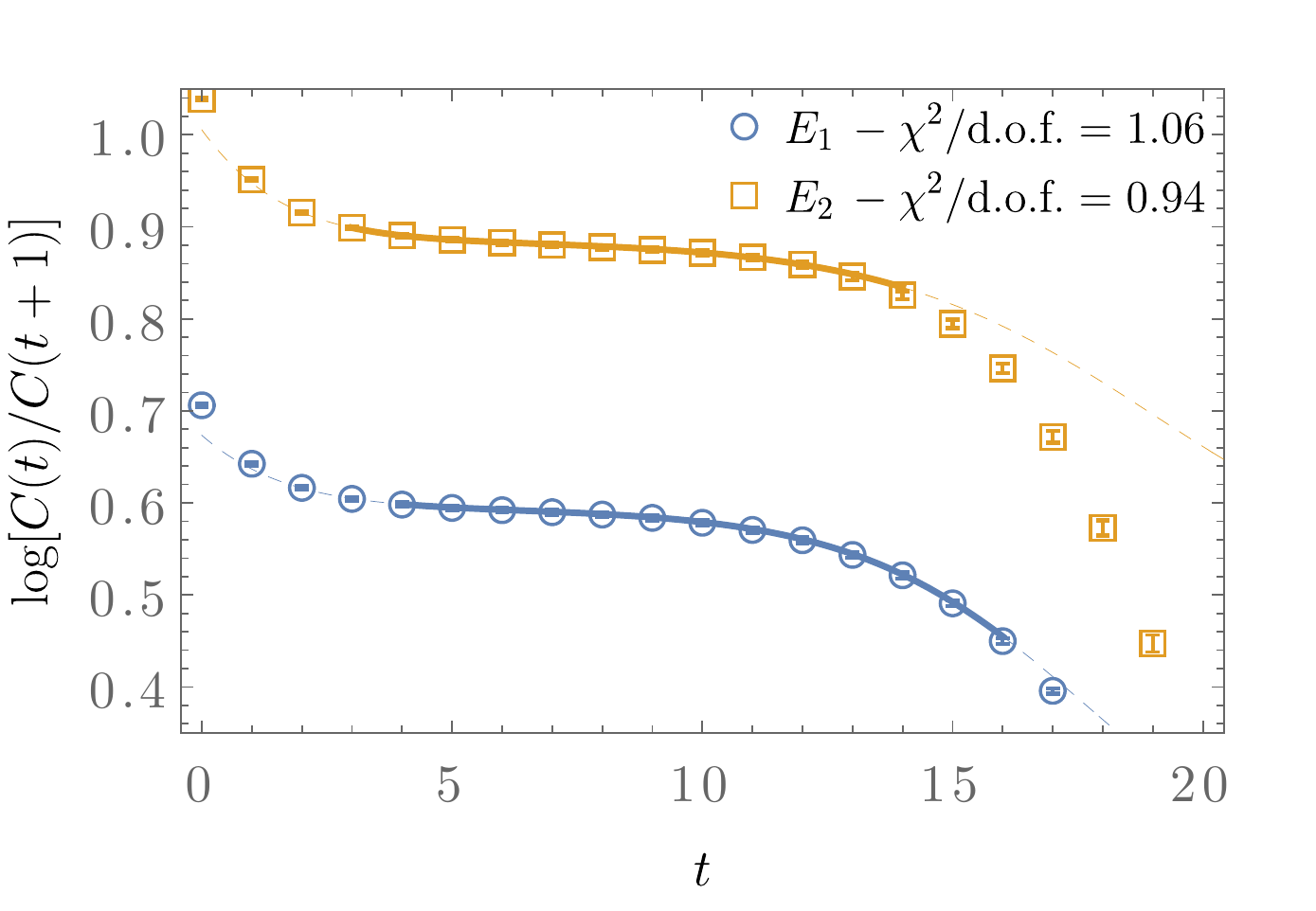}
    \caption{Effective masses of the ground ($E_1$) and first excited state ($E_2$) correlation functions from solving the GEVP on ensemble $\mathcal{E}_1$ in irrep $A_{1u}$.  The solid line represents the best fit plotted in the fitting range that minimizes the $\chi^2$.  The dashed line is the extension of the best fit beyond the fitting range.}
    \label{fig:corr_fit}
\end{figure}

\begin{table}
\centering
    \begin{tabular*}{0.6\columnwidth}{C @{\extracolsep{\fill}}C C C C C C}
        \hline\hline
        n &     a_1 & b_1               & a_2 & b_2                 & a_3 & b_3 \\
        \hline
        1 &     0.151(4) & 6.6(2)       & 0.810(4) & 3.074(3)       & 0.00029(3) & 0.822(3) \\
        2 &     0.211(3) & 8.4(1)       & 0.752(4) & 4.579(9)       & 0.005(4)   & 2.6(3)\\
        \hline\hline
    \end{tabular*}
    \caption{Fit parameters from the fits shown in Fig.~\ref{fig:corr_fit} using a functional form as described in the text.  The $a_i$ are coefficients of decaying exponentials with rates $b_i$. The coefficient $b_2$ is the energy of the state of interest, $b_1$ is the excited state contamination, and $b_3$ is $\delta E$ appearing from ``temporal wraparound".}  
    \label{table:fit_params}
\end{table}

\section{Lattice Energy Eigenvalues and Predicted Spectrum}
\label{appendix:energies}

\begin{table}[H]
  \centering
  \caption{\label{table:energies}
  Energies extracted as discussed in the main text.  $\Gamma$ is the irrep, $E/m_{\pi}$ is the energy eigenvalue in units of the pion mass.  The TRUF predictions are made using the low-energy constants from Refs.~\cite{Gasser:1984gg} and \cite{Mai:2019fba}, denoted by GL and GW, respectively.  Jackknife samples for all lattice energies are provided with the arxiv submission.}

  \def\arraystretch{1.35}
  \begin{tabular*}{0.6\columnwidth}{L @{\extracolsep{\fill}}L L C C}
  \hline\hline
    \text{Ensemble}    &   \Gamma    &   E/m_{\pi}     &   \text{TRUF[GL]} &   \text{TRUF[GW]}\\
    \hline
    \mathcal{E}_1         &   A_{1u}       &   3.074(3)    & 3.067 & 3.080\\
    &               &   4.579(9)   & 4.584 & 4.568\\
    &   E_u         &   4.469(8)    & 4.455 & 4.457\\
    \hline
    \mathcal{E}_2         &   A_{1u}       &   3.060(5)    & 3.059  & 3.059\\
    &               &   4.041(8)    & 4.045  & 4.052\\
    &               &   4.476(7)     & 4.486  & 4.486\\
    &   B_{1u}       &   4.412(8)    & 4.420  & 4.420 \\
    \hline
    \mathcal{E}_3         &   A_{1u}       &   3.023(5)    & 3.034  & 3.034\\
    &               &   3.479(5)    & 3.489  & 3.494\\
    &               &   4.158(5)    & 4.150  & 4.152\\
    &               &   4.407(6)    & 4.408  & 4.412\\
    &               &   4.441(6)    & 4.460  & 4.458\\
    &               &   4.738(6)    & 4.755  & 4.740\\
    &               &   4.833(6)    & 4.845  & 4.844\\
    &   A_{2g}       &   4.159(6)    & 4.152  & 4.152\\
    &               &   4.797(6)    & 4.808  & 4.794\\
    &   B_{1u}       &   4.402(5)    & 4.406  & 4.408\\
    &               &   4.662(5)    & 4.662  & 4.663\\
    &               &   4.795(7)    & 4.781  & 4.781\\
    &   B_{2u}       &   4.729(6)    & 4.725  & 4.725 \\
    &   E_g         &  4.761(6)    & 4.764  & 4.764\\
    &   E_u         &  4.661(5)    & 4.659  & 4.659\\
    &               &  4.733(7)    & 4.735  & 4.740\\

    \hline
    \mathcal{E}_4       &   A_{1u}   &      3.145(10)    & 3.127    & 3.133\\
    \mathcal{E}_4^{\left[001\right]} & A_2     &    3.959(45)     & 3.908    & 3.908\\
    \hline
    \mathcal{E}_5       &   A_{1u}   &   3.106(9)      &  3.110    & 3.114\\
    \mathcal{E}_5^{\left[001\right]}   &   A_2     &  3.772(12)     & 3.745     & 3.750 \\
    \hline
    \mathcal{E}_6   &     A_{1u}     &  3.098(8)      & 3.093     & 3.097\\
    &               &  4.676(14)      & 4.654     & 4.651\\
    \mathcal{E}_6^{\left[001\right]} &   A_2   &  3.620(8)     & 3.616    & 3.622\\
    \hline\hline
  \end{tabular*}

\end{table}

\clearpage

\section{List of operators used}
\label{appendix:operators}
Here we collect all of the operators that were used to extract the finite volume spectrum.  The operators were constructed as described in the main text using equation~2 in the main text. Note that the overall normalization of the individual operators is arbitrary.

\begin{table}[!h]
  \caption{Three pion interpolating operators used for ensemble ${\cal E}_1$ with total momentum $\*P=0$, transforming according to irrep $\Upgamma$. The label $n$ indicates the order of the non-interacting energy levels with 1 being the lowest energy level in that irrep.  Where relevant, irrep row is indicated by an integer superscript. Operators are denoted in terms of the constituent momenta as $[\*d_1][\*d_2][\*d_3]$, where $\*p=\frac{2\pi}{L}\*d$.  Empty entries indicate vanishing coefficients.}
  \label{tab:ops1}
  \def\arraystretch{1.2}
  \begin{tabular*}{0.5\columnwidth}{C C |@{\extracolsep{\fill}}C C C C C C C C C C C}
  \hline\hline
    \Upgamma & n & \rot{[\,0\,\,0\,\,0\,][\,0\,\,0\,\,0\,][\,0\,\,0\,\,0\,]} &
    \rot{[-1\,\,0\,\,0\,][\,0\,\,0\,\,0\,][\,1\,\,0\,\,0\,]} &
    \rot{[\,0\,-1\,\,0\,][\,0\,\,0\,\,0\,][\,0\,\,1\,\,0\,]} &
    \rot{[\,0\,\,0\,-1\,][\,0\,\,0\,\,0\,][\,0\,\,0\,\,1\,]} &
    \rot{[-1\,-1\,\,0\,][\,0\,\,0\,\,0\,][\,1\,\,1\,\,0\,]} &
    \rot{[-1\,\,0\,-1\,][\,0\,\,0\,\,0\,][\,1\,\,0\,\,1\,]} &
    \rot{[-1\,\,0\,\,1\,][\,0\,\,0\,\,0\,][\,1\,\,0\,-1\,]} &
    \rot{[-1\,\,1\,\,0\,][\,0\,\,0\,\,0\,][\,1\,-1\,\,0\,]} &
    \rot{[\,0\,-1\,-1\,][\,0\,\,0\,\,0\,][\,0\,\,1\,\,1\,]} &
    \rot{[\,0\,-1\,\,1\,][\,0\,\,0\,\,0\,][\,0\,\,1\,-1\,]\,\,} & \phantom{gho} \\  
    \hline 
    A_{1u} & 1 & 1 & & & & & & & & &\\
    & 2 & & 1 & 1 & 1 & & & & & &\\ 
    & 3 & & & & & 1 & 1 & 1 & 1 & 1 & 1\\
    E_u^{(1)} & 1 & & 1 & -1 & & & & & & &\\ 
    & 2 & & & & & & 1 & 1 & & -1 & -1\\
    E_u^{(2)} & 1 & & 1 & 1 & -2 & & & & & &\\ 
    & 2 & & & & & -2 & 1 & 1 & -2 & 1 & 1 \\
    \hline\hline
  \end{tabular*}
  %
\vspace{1cm}
  \centering
  \caption{Same as Table~\ref{tab:ops1} but for ensemble ${\cal E}_2$ with total momentum $\*P=0$.}
  \label{tab:ops2}
  \def\arraystretch{1.2}
  \begin{tabular*}{0.4\columnwidth}{C C |@{\extracolsep{\fill}}C C C C C C C C C}
  \hline\hline
    \Upgamma & n & \rot{[\,0\,\,0\,\,0\,][\,0\,\,0\,\,0\,][\,0\,\,0\,\,0\,]} &
    \rot{[\,0\,\,0\,-1\,][\,0\,\,0\,\,0\,][\,0\,\,0\,\,1\,]} &
    \rot{[-1\,\,0\,\,0\,][\,0\,\,0\,\,0\,][\,1\,\,0\,\,0\,]} &
    \rot{[\,0\,-1\,\,0\,][\,0\,\,0\,\,0\,][\,0\,\,1\,\,0\,]} &
    \rot{[-1\,\,0\,-1\,][\,0\,\,0\,\,0\,][\,1\,\,0\,\,1\,]} &
    \rot{[-1\,\,0\,\,1\,][\,0\,\,0\,\,0\,][\,1\,\,0\,-1\,]} &
    \rot{[\,0\,-1\,-1\,][\,0\,\,0\,\,0\,][\,0\,\,1\,\,1\,]} &
    \rot{[\,0\,-1\,\,1\,][\,0\,\,0\,\,0\,][\,0\,\,1\,-1\,]\,\,} & \phantom{gho} \\  
    \hline 
    A_{1u} & 1 & 1 & & & & & & &\\ 
    & 2 & & 1 & & & & & &\\
    & 3 & & & 1 & 1 & & & &\\ 
    & 4 & & & & & 1 & 1 & 1 & 1\\
    B_{1u} & 1 & & & 1 & -1 & & & &\\ 
    & 2 & & & & & 1 & 1 & -1 & -1\\
    \hline\hline
  \end{tabular*}
\end{table}

\begin{table}[!h]
  \caption{Same as Table~\ref{tab:ops1} but for ensemble ${\cal E}_3$ with total momentum $\*P=0$. {\it Part 1}.}
  \label{tab:ops3a}

  \def\arraystretch{1.2}
  \begin{tabular*}{0.65\columnwidth}{C C |@{\extracolsep{\fill}}C C C C C C C C C C C C C C}
  \hline\hline
    \Upgamma & n & \rot{[\,0\,\,0\,\,0\,][\,0\,\,0\,\,0\,][\,0\,\,0\,\,0\,]} &
    \rot{[\,0\,\,0\,-1\,][\,0\,\,0\,\,0\,][\,0\,\,0\,\,1\,]} &
    \rot{[\,0\,\,0\,-2\,][\,0\,\,0\,\,1\,][\,0\,\,0\,\,1\,]} &
    \rot{[\,0\,\,0\,-1\,][\,0\,\,0\,-1\,][\,0\,\,0\,\,2\,]} &
    \rot{[-1\,\,0\,\,0\,][\,0\,\,0\,\,0\,][\,1\,\,0\,\,0\,]} &
    \rot{[\,0\,-1\,\,0\,][\,0\,\,0\,\,0\,][\,0\,\,1\,\,0\,]} &
    \rot{[\,0\,\,0\,-2\,][\,0\,\,0\,\,0\,][\,0\,\,0\,\,2\,]} &
    \rot{[-1\,\,0\,-1\,][\,0\,\,0\,\,0\,][\,1\,\,0\,\,1\,]} &
    \rot{[-1\,\,0\,\,1\,][\,0\,\,0\,\,0\,][\,1\,\,0\,-1\,]} &
    \rot{[\,0\,-1\,-1\,][\,0\,\,0\,\,0\,][\,0\,\,1\,\,1\,]} &
    \rot{[\,0\,-1\,\,1\,][\,0\,\,0\,\,0\,][\,0\,\,1\,-1\,]} &
    \rot{[\,0\,\,0\,-3\,][\,0\,\,0\,\,1\,][\,0\,\,0\,\,2\,]} &
    \rot{[\,0\,\,0\,-2\,][\,0\,\,0\,-1\,][\,0\,\,0\,\,3\,]\,\,} & \phantom{gho} \\  
    \hline 
    A_{1u} & 1 & 1 & & & & & & & & & & & &\\ 
    & 2 & & 1 & & & & & & & & & & &\\ 
    & 3 & & & 1 & 1 & & & & & & & & &\\ 
    & 4 & & & & & 1 & 1 & & & & & & &\\ 
    & 5 & & & & & & & 1 & & & & & &\\ 
    & 6 & & & & & & & & 1 & 1 & 1 & 1 & &\\ 
    & 8 & & & & & & & & & & & & 1 & 1\\
    A_{2g} & 1 & & & 1 & -1 & & & & & & & & &\\ 
    & 3 & & & & & & & & & & & & 1 & -1\\
    E_u^{(2)} & 1 & & & & & & & & 1 & -1 & & & &\\ 
    B_{1u} & 1 & & & & & 1 & -1 & & & & & & &\\ 
    & 2 & & & & & & & & 1 & 1 & -1 & -1 & &\\ 
\hline\hline
  \end{tabular*}


\vspace{1cm}
  \caption{Same as Table~\ref{tab:ops1} but for ensemble ${\cal E}_3$ with total momentum $\*P=0$. {\it Part 2}.}
  \label{tab:ops3b}
  
  \def\arraystretch{1.2}
  \begin{tabular*}{0.8\columnwidth}{C C |@{\extracolsep{\fill}}C C C C C C C C C C C C C C C C C}
  \hline\hline
    \Upgamma & n & \rot{[-1\,\,0\,-1\,][\,0\,\,0\,\,1\,][\,1\,\,0\,\,0\,]} &
    \rot{[-1\,\,0\,\,0\,][\,0\,\,0\,-1\,][\,1\,\,0\,\,1\,]} &
    \rot{[-1\,\,0\,\,0\,][\,0\,\,0\,\,1\,][\,1\,\,0\,-1\,]} &
    \rot{[-1\,\,0\,\,1\,][\,0\,\,0\,-1\,][\,1\,\,0\,\,0\,]} &
    \rot{[\,0\,-1\,-1\,][\,0\,\,0\,\,1\,][\,0\,\,1\,\,0\,]} &
    \rot{[\,0\,-1\,\,0\,][\,0\,\,0\,-1\,][\,0\,\,1\,\,1\,]} &
    \rot{[\,0\,-1\,\,0\,][\,0\,\,0\,\,1\,][\,0\,\,1\,-1\,]} &
    \rot{[\,0\,-1\,\,1\,][\,0\,\,0\,-1\,][\,0\,\,1\,\,0\,]} &
    \rot{[-1\,\,0\,-2\,][\,0\,\,0\,\,1\,][\,1\,\,0\,\,1\,]} &
    \rot{[-1\,\,0\,-1\,][\,0\,\,0\,-1\,][\,1\,\,0\,\,2\,]} &
    \rot{[-1\,\,0\,\,1\,][\,0\,\,0\,\,1\,][\,1\,\,0\,-2\,]} &
    \rot{[-1\,\,0\,\,2\,][\,0\,\,0\,-1\,][\,1\,\,0\,-1\,]} &
    \rot{[\,0\,-1\,-2\,][\,0\,\,0\,\,1\,][\,0\,\,1\,\,1\,]} &
    \rot{[\,0\,-1\,-1\,][\,0\,\,0\,-1\,][\,0\,\,1\,\,2\,]} &
    \rot{[\,0\,-1\,\,1\,][\,0\,\,0\,\,1\,][\,0\,\,1\,-2\,]} &
    \rot{[\,0\,-1\,\,2\,][\,0\,\,0\,-1\,][\,0\,\,1\,-1\,]\,\,} & \phantom{gho}\\  
    \hline 
    A_{1u} & 7 & 1 & 1 & 1 & 1 & 1 & 1 & 1 & 1 & & & & & & & &\\ 
    & 9 & & & & & & & & & 1 & 1 & 1 & 1 & 1 & 1 & 1 & 1\\
    A_{2g} & 2 & 1 & -1 & 1 & -1 & 1 & -1 & 1 & -1 & & & & & & & &\\ 
    & 4 & & & & & & & & & 1 & -1 & 1 & -1 & 1 & -1 & 1 & -1\\
    E_u^{(2)} & 2 & 1 & 1 & -1 & -1 & & & & & & & & & & & &\\ 
    & 3 & & & & & & & & & 1 & 1 & -1 & -1 & & & &\\ 
    E_g^{(1)} & 1 & 1 & -1 & -1 & 1 & & & & & & & & & & & &\\ 
    & 2 & & & & & & & & & 1 & -1 & -1 & 1 & & & &\\ 
    B_{1u} & 3 & 1 & 1 & 1 & 1 & -1 & -1 & -1 & -1 & & & & & & & &\\ 
    & 4 & & & & & & & & & 1 & 1 & 1 & 1 & -1 & -1 & -1 & -1\\
    B_{2u} & 1 & 1 & -1 & 1 & -1 & -1 & 1 & -1 & 1 & & & & & & & &\\ 
    & 2 & & & & & & & & & 1 & -1 & 1 & -1 & -1 & 1 & -1 & 1\\
    \hline\hline
  \end{tabular*}

\end{table}

\begin{table}[!h]
  \caption{Same as Table~\ref{tab:ops1} but for ensembles ${\cal E}_4$, ${\cal E}_5$, and ${\cal E}_6$ with total momentum $\*P=0$.}
  \label{tab:ops456}

  \def\arraystretch{1.2}
  \begin{tabular*}{0.5\columnwidth}{C C |@{\extracolsep{\fill}}C C C C C C C C C C C}
  \hline\hline
    \Upgamma & n & \rot{[\,0\,\,0\,\,0\,][\,0\,\,0\,\,0\,][\,0\,\,0\,\,0\,]} &
    \rot{[-1\,\,0\,\,0\,][\,0\,\,0\,\,0\,][\,1\,\,0\,\,0\,]} &
    \rot{[\,0\,-1\,\,0\,][\,0\,\,0\,\,0\,][\,0\,\,1\,\,0\,]} &
    \rot{[\,0\,\,0\,-1\,][\,0\,\,0\,\,0\,][\,0\,\,0\,\,1\,]} &
    \rot{[-1\,-1\,\,0\,][\,0\,\,0\,\,0\,][\,1\,\,1\,\,0\,]} &
    \rot{[-1\,\,0\,-1\,][\,0\,\,0\,\,0\,][\,1\,\,0\,\,1\,]} &
    \rot{[-1\,\,0\,\,1\,][\,0\,\,0\,\,0\,][\,1\,\,0\,-1\,]} &
    \rot{[-1\,\,1\,\,0\,][\,0\,\,0\,\,0\,][\,1\,-1\,\,0\,]} &
    \rot{[\,0\,-1\,-1\,][\,0\,\,0\,\,0\,][\,0\,\,1\,\,1\,]} &
    \rot{[\,0\,-1\,\,1\,][\,0\,\,0\,\,0\,][\,0\,\,1\,-1\,]\,\,} & \phantom{gho}\\  
    \hline 
    A_{1u} & 1 & 1 & & & & & & & & &\\ 
    & 2 & & 1 & 1 & 1 & & & & & &\\ 
    & 3 & & & & & 1 & 1 & 1 & 1 & 1 & 1\\
    \hline\hline
  \end{tabular*}
\vspace{1cm}
  \caption{Same as Table~\ref{tab:ops1} but for ensembles ${\cal E}_4$, ${\cal E}_5$, and ${\cal E}_6$ with total momentum $\*P=\frac{2\pi}{L}[001]$ in lattice units.}
  \label{tab:ops456boost}

  \def\arraystretch{1.2}
  \begin{tabular*}{0.4\columnwidth}{C C |@{\extracolsep{\fill}}C C C C C C C C C}
  \hline\hline
    \Upgamma & n & \rot{[\,0\,\,0\,\,0\,][\,0\,\,0\,\,0\,][\,0\,\,0\,\,1\,]} &
    \rot{[\,0\,\,0\,-1\,][\,0\,\,0\,\,1\,][\,0\,\,0\,\,1\,]} &
    \rot{[-1\,\,0\,\,0\,][\,0\,\,0\,\,0\,][\,1\,\,0\,\,1\,]} &
    \rot{[-1\,\,0\,\,1\,][\,0\,\,0\,\,0\,][\,1\,\,0\,\,0\,]} &
    \rot{[\,0\,-1\,\,0\,][\,0\,\,0\,\,0\,][\,0\,\,1\,\,1\,]} &
    \rot{[\,0\,-1\,\,1\,][\,0\,\,0\,\,0\,][\,0\,\,1\,\,0\,]} &
    \rot{[-1\,\,0\,\,0\,][\,0\,\,0\,\,1\,][\,1\,\,0\,\,0\,]} &
    \rot{[\,0\,-1\,\,0\,][\,0\,\,0\,\,1\,][\,0\,\,1\,\,0\,]\,\,} & \phantom{gho}\\  
    \hline 
    A_2 & 1 & 1 & & & & & & &\\ 
    & 2 & & 1 & & & & & &\\ 
    & 3 & & & 1 & 1 & 1 & 1 & &\\ 
    & 4 & & & & & & & 1 & 1\\
    \hline\hline
  \end{tabular*}
\end{table}
\vfill
\end{widetext}

\end{document}